\documentclass[twocolumn,journal]{IEEEtran}
\usepackage[T1]{fontenc}
\usepackage[latin9]{inputenc}
\usepackage{verbatim}
\usepackage{float}
\usepackage{amsmath}
\usepackage{amsthm}
\usepackage{amssymb}
\usepackage{graphicx}
\usepackage{color}
\usepackage[unicode=true,
 bookmarks=true,bookmarksnumbered=true,bookmarksopen=true,bookmarksopenlevel=1,
 breaklinks=false,pdfborder={0 0 1},backref=false,colorlinks=false]
 {hyperref}
\hypersetup{pdftitle={Your Title},
 pdfauthor={Your Name},
 pdfborderstyle=,pdfpagelayout=OneColumn,pdfnewwindow=true,pdfstartview=XYZ,plainpages=false}

\makeatletter

\floatstyle{ruled}
\newfloat{algorithm}{tbp}{loa}
\providecommand{\algorithmname}{Algorithm}
\floatname{algorithm}{\protect\algorithmname}

\theoremstyle{plain}
\newtheorem{thm}{\protect\theoremname}
\theoremstyle{remark}
\newtheorem{rem}[thm]{\protect\remarkname}

\usepackage{algorithmic}
\floatname{algorithm}{Algorithm}
\usepackage{cite}
\usepackage{siunitx}
\usepackage{acronym}

\AtBeginDocument{\acrodef{AP}{access point}
\acrodef{CPU}{central processing unit}
\acrodef{CSI}{channel state information}
\acrodef{MIMO}{multiple\mbox{-}input multiple\mbox{-}output}
\acrodef{MISO}{multiple\mbox{-}input single\mbox{-}output}
\acrodef{MMSE}{minimum mean square error}
\acrodef{L-MMSE}{local minimum mean square error}
\acrodef{SNR}{signal-to-noise ratio}
\acrodef{SINR}{signal-to-interference plus noise ratio}
\acrodef{SE}{spectral efficiency}
\acrodef{EE}{energy efficiency}
\acrodef{SEmax}{spectral efficiency maximization}
\acrodef{EEmax}{energy efficiency maximization}
\acrodef{MR}{maximum ratio}
\acrodef{GP}{geometric programming}
\acrodef{NOMA}{non-orthogonal multiple access}
\acrodef{QoS}{quality of service}
\acrodef{MRC}{maximum-ratio combining}
\acrodef{APG}{accelerated projected gradient}
\acrodef{CDF}{cumulative distribution function}
\acrodef{AO}{alternating optimization}
\acrodef{5G}{fifth generation}
\acrodef{ZF}{zero-forcing}
\acrodef{MD}{mirror descent}
\acrodef{MP}{mirror prox}
\acrodef{EVP}{eigenvalue problem}
\acrodef{GDA}{gradient descent accent}
\acrodef{LP}{linear programming}}

\DeclareMathOperator{\maximize}{maximize}
\DeclareMathOperator{\minimize}{minimize}
\DeclareMathOperator{\st}{subject \, to}

\newcommand{\herm}{^{\mbox{\scriptsize H}}}
\newcommand{\trans}{^{\mbox{\scriptsize T}}}
\makeatother

\providecommand{\remarkname}{Remark}
\providecommand{\theoremname}{Theorem}
\begin{document}
\title{Mirror Prox Algorithm for Large-Scale Cell-Free Massive MIMO Uplink
Power Control}
\author{Muhammad~Farooq, Hien~Quoc~Ngo, and~Le~Nam~Tran\thanks{M.~Farooq and L.-N. Tran are with School of Electrical and Electronic Engineering, University College Dublin, Ireland. Email: muhammad.farooq@ucdconnect.ie; nam.tran@ucd.ie}\thanks{H. Q. Ngo is with Institute of Electronics, Communications and Information Technology, Queen\textquoteright s University Belfast, Belfast BT3 9DT, U.K. Email: hien.ngo@qub.ac.uk}
\vspace{-1mm}}
\maketitle
\begin{abstract}
We consider the problem of max-min fairness for uplink cell-free massive multiple-input multiple-output (MIMO) subject to per-user power constraints. The standard framework for solving the considered problem is to separately solve two subproblems: the receiver filter coefficient design and the power control problem. While the former has a closed-form solution, the latter has been solved using either second-order methods of high computational complexity or a first-order method that provides an approximate solution. To deal with these drawbacks of the existing methods, we propose a mirror prox based method for the power control problem by equivalently reformulating it as a convex-concave problem and applying the mirror prox algorithm to find a saddle point. The simulation results establish the optimality of the proposed solution and demonstrate that it is more efficient than the known methods. We also conclude that for large-scale cell-free massive MIMO, joint optimization of linear receive combining and power control provides significantly better user fairness than the power control only scheme in which receiver coefficients are fixed to unity.
\end{abstract}


\vspace{-1mm}
\section{Introduction}

\bstctlcite{IEEEexample:BSTcontrol} The recently evolved form of
the massive \ac{MIMO} for the beyond \ac{5G} networks is cell-free
massive \ac{MIMO} \cite{Ngo2017cfmm}, in which one or several
central processing units (CPUs) control a large number of low-cost
and low-power \acp{AP} to serve a large number of users. The \acp{AP}
and users are distributed in a large coverage area. Owing to the high
array gain, multiplexing gain, and macro-diversity gain, cell-free
massive MIMO offers many advantages such as high energy efficiency,
uniform quality of service, and flexible and cost-effective deployment
\cite{Zhang2019access}.

To achieve the above benefits, suitable power allocation algorithms
need to be employed to control the near/far effect. Since the numbers
of APs and users are large, the resulting power control problems consist
of many variables. Thus, simple and low-complexity solutions to power
control optimization problems for cell-free massive \ac{MIMO} are of
particular interest. In \cite{Ngo2017cfmm}, the problem of maximization
of the minimum uplink rate (a.k.a. \emph{max-min fairness}) for single-antenna
\acp{AP} and single-antenna users was solved using a bisection
algorithm in combination with \ac{LP}, where \ac{MRC} was considered
at each \ac{AP} for signal detection. In \cite{Bashar2019maxmin}, Bashar \emph{et al.} solved this max-min fairness problem by alternatively solving two subproblems: (i) the receiver coefficient design which is in fact a generalized
\ac{EVP}, and (ii) the power control problem which was solved using \ac{GP}. A similar \ac{AO}-based scheme was used in \cite{Bashar2018mixQoS} to solve a mixed \ac{QoS} problem including the max-min fairness for a set of users and a fixed \ac{QoS} for the remaining users. Again, the power control subproblem
was solved using \ac{GP}. In  \cite{Mai2019uplinkMU}, Mai \emph{et al.} employed the \ac{ZF} technique to detect symbols from multi-antenna users where the power
control problem was solved using the bisection algorithm. A common feature of all aforementioned studies is the use of \ac{LP} or (more complex) \ac{GP} by means of off-the-shelf convex solvers, and thus are suitable for small-scale scenarios. In particular, a low complexity first-order \ac{APG} algorithm was applied to the max-min fairness problem following Nesterov's smoothing technique \cite{Nesterov2005a} in both downlink \cite{Farooq2020,Farooq2021} and uplink \cite{farooq2020APG} channels. However, this method can only produce an approximate solution whose accuracy is inversely proportional to the number of users, and hence, is not suitable for very large numbers of users. This clearly calls for novel solutions to the power control problem, which are more computationally  efficient than LP or GP methods and more accurate than the APG method.

In this paper, we consider the max-min fairness problem subject to the power constraint at each user. We utilize the \ac{MRC} technique to detect users' signals using the channel estimates. Similar to the known approaches, we also split the considered problem into two subproblems which are alternately solved. In particular, for the power control subproblem, we propose a \ac{MP}-based algorithm. To achieve this, we first reformulate the \emph{nonconvex} power control problem into an equivalent convex-concave program and then propose an \ac{MP} method \cite{nemirovski2004prox} to find a saddle point. The proposed method only requires the first order information of the objective, and thus, is shown to be much faster than the LP-based and GP-based methods. Moreover, it outperforms other existing first-order  methods in terms of achievable rate performance.

\emph{Notation}s: Bold lower and upper case letters represent vectors
and matrices. $\mathcal{CN}(\boldsymbol{0},\mathbf{K})$ denotes the
multivariate circularly symmetric complex Gaussian random distribution
with zero mean and co-variance matrix $\mathbf{K}$. $\mathbf{X}\trans$
and $\mathbf{X}\herm$ stand for the transpose and Hermitian of $\mathbf{X}$,
respectively. Notation $\mathbf{e}_{i}$ denotes the $i$-th column
of the identity matrix. $\nabla f(\mathbf{x})$ represents the gradient
of $f(\mathbf{x})$. $\left\Vert \cdot\right\Vert $ denotes the Euclidean
or $\ell_{2}$-norm and $|\cdot|$ is the absolute value of the argument.
$\mathcal{P}_{\mathcal{C}}(\mathbf{x})$ denotes the projection of
$\mathbf{x}$ onto the convex set $\mathcal{C}$, i.e. $\mathcal{P}_{\mathcal{C}}(\mathbf{x})=\underset{\mathbf{z}\in\mathcal{C}}{\arg\min}\bigl\Vert\mathbf{x}-\mathbf{z}\bigr\Vert$.

\vspace{-1mm}
\section{System Model and Problem Formulation}

Consider a cell-free massive \ac{MIMO} uplink scenario where $L$ single-antenna users are served by $M$ \acp{AP} with $K$ antennas per \ac{AP}. All APs and users are randomly distributed in a coverage area. The APs are connected to a CPU via high-capacity dedicated links. We denote by $\zeta_{ml}$ the large-scale fading coefficient between the $l$-th user and the $m$-th \ac{AP}, and $\tilde{\mathbf{h}}_{ml}\sim\mathcal{CN}(\boldsymbol{0},\textbf{I}_{K})$
is the vector of the small-scale fading coefficient for all antennas at the $m$-th \ac{AP}. The channel vector between the $l$-th user and the $m$-th \ac{AP} is modeled as  \begin{equation}
\mathbf{h}_{ml}=\zeta_{ml}^{1/2}\tilde{\mathbf{h}}_{ml}.\label{eq:channel}
\end{equation}
The length (in symbols) of the  uplink training is denoted by $\tau_{p}$.
Let $\eta_{p}$ be the transmit power of each pilot symbol and $\boldsymbol{\phi}_{l}$ be the pilot sequence of unit norm transmitted from user $l$ to the \acp{AP}. Let $\hat{\mathbf{h}}_{ml}$ be the \ac{MMSE} channel
estimate of $\mathbf{h}_{ml}$. Then $\hat{\mathbf{h}}_{ml}\sim\mathcal{CN}(\boldsymbol{0},g_{ml}\textbf{I}_{K})$,
where \cite{Ngo2017cfmm} 
\begin{equation}
g_{ml}\triangleq\frac{\eta_{p}\tau_{p}\zeta_{ml}^{2}}{\eta_{p}\tau_{p}\sum_{i=1}^{L}\zeta_{mi}\left|\boldsymbol{\phi}_{i}\herm\boldsymbol{\phi}_{l}\right|^{2}+1}.
\end{equation}
\begin{comment}
Here you need to tell the setup before stating the problem. Define
the power and then write down the SINR and the achievable rate first.
\end{comment}
To formulate the problem of interest, we define $\mathbf{p}=[p_{1},\ldots,p_{L}]$
as the vector of power control coefficients, $\mathbf{q}=[\mathbf{q}_{1};\mathbf{q}_{2};\ldots;\mathbf{q}_{L}]$
as the receiver coefficient where $\mathbf{q}_{l}=[q_{1l};q_{2l};\ldots;q_{Ml}]$
associates with the $l$-th user, and $\bar{p}$ as the maximum transmit
power at each individual user. With this setup, an achievable \ac{SINR}
is given by \cite{Ngo2017cfmm}
\begin{equation}
\gamma_{l}(\mathbf{p},\mathbf{q})=\frac{\mathbf{q}_{l}\herm\bigl(\mathbf{g}_{ll}\mathbf{g}_{ll}\herm p_{l}\bigr)\mathbf{q}_{l}}{\mathbf{q}_{l}\herm\bigl(\sum_{i\neq l}^{L}\mathbf{g}_{li}\mathbf{g}_{li}\herm p_{i}+\frac{1}{K}\sum_{i=1}^{L}\bar{\mathbf{G}}_{li}p_{i}+\frac{1}{K}\tilde{\mathbf{G}}_{l}\bigr)\mathbf{q}_{l}}.\label{eq:SINR}
\end{equation}
In \eqref{eq:SINR}, $\bar{\mathbf{G}}_{li}\in\mathbb{R}_{+}^{M\times M}$
and $\tilde{\mathbf{G}}_{l}\in\mathbb{R}_{+}^{M\times M}$ are diagonal
matrices with $[\bar{\mathbf{G}}_{li}]_{m,m}=g_{ml}\zeta_{mi}$ and
$[\tilde{\mathbf{G}}_{l}]_{m,m}=g_{ml}$, respectively, and 
\begin{equation}
\mathbf{g}_{li}\triangleq\left|\boldsymbol{\phi}_{l}\herm\boldsymbol{\phi}_{i}\right|\left[g_{1l}\frac{\zeta_{1i}}{\zeta_{1l}};g_{2l}\frac{\zeta_{2i}}{\zeta_{2l}};\ldots;g_{Ml}\frac{\zeta_{Mi}}{\zeta_{Ml}}\right].
\end{equation}
Note that the achievable rate of the $l$-th user is given by 
\begin{equation}
\mathcal{R}_{l}(\mathbf{p},\mathbf{q})=\log_{2}\left(1+\gamma_{l}(\mathbf{p},\mathbf{q})\right).\label{eq:SEk-1}
\end{equation}

Similar to \cite{Bashar2019maxmin}, we consider the problem of maximization
of the minimum rate, which is mathematically stated as
\begin{equation}
\begin{array}{ll}
\underset{\mathbf{q},\mathbf{p}}{\maximize} & \quad\underset{1\leq l\leq L}{\min}\gamma_{l}(\mathbf{p},\mathbf{q})\\
\st & \quad\left\Vert \mathbf{q}_{l}\right\Vert =1,\,l=1,\ldots,L,\\
 & \quad0\leq\mathbf{p}\leq\bar{p}.
\end{array}\tag{\ensuremath{\mathcal{P}_{1}}}\label{eq:equivProb}
\end{equation}
First, we note that \eqref{eq:equivProb} is nonconvex, and thus, finding a globally optimal solution is difficult and not practically useful. The standard framework for solving \eqref{eq:equivProb} which is adopted in previous studies such as \cite{Ngo2017cfmm,Bashar2019maxmin,Bashar2018mixQoS} is based on \ac{AO}. In this way, two subproblems arise: the receiver coefficient design and the power control problem. Specifically, the receiver coefficient design is obtained by fixing the power allocation $\mathbf{p}$, which admits a closed-form solution given by \cite{Bashar2019maxmin,Bashar2018mixQoS}
\begin{equation}
\mathbf{q}_{l}^{*}=\frac{\sqrt{p_{l}}\mathbf{W}_{l}^{-1}\mathbf{g}_{ll}}{\bigl\Vert\sqrt{p_{l}}\mathbf{W}_{l}^{-1}\mathbf{g}_{ll}\bigr\Vert},l=1,2,\ldots,L,\label{eq:optimalU}
\end{equation}
where $\mathbf{W}_{l}=\sum_{i\neq l}^{L}\mathbf{g}_{li}\mathbf{g}_{li}\herm p_{i}+\frac{1}{K}\sum_{i=1}^{L}\bar{\mathbf{G}}_{li}p_{i}+\frac{1}{K}\tilde{\mathbf{G}}_{l}$.
After updating the receiver coefficients for all users, we need to find the power coefficients, leading to the following power control problem: 
\begin{equation}
\boxed{
%\begin{aligned}\underset{\mathbf{p}}{\maximize} & \quad\underset{1\leq l\leq L}{\min}\gamma_{l}(\mathbf{p})\\ 
%\st & \quad0\leq\mathbf{p}\leq\bar{p}.
%\end{aligned}
\underset{0\leq\mathbf{p}\leq\bar{p}}{\maximize} \quad\underset{1\leq l\leq L}{\min}\gamma_{l}(\mathbf{p})
\tag{\ensuremath{\mathcal{P}_{2}}}}\label{eq:powerAlloc}
\end{equation}

To solve \eqref{eq:powerAlloc}, existing methods \cite{Ngo2017cfmm,Bashar2019maxmin,Bashar2018mixQoS} involve second order methods, i.e., \ac{GP} or \ac{LP}, which require very high complexity, and thus, is not suitable for large-scale cell-free massive MIMO. To overcome the complexity issue, a first-order algorithm was introduced in \cite{farooq2020APG} but it cannot yield a solution of high accuracy for a large number of users. In the next section, we propose a more efficient method for solving \eqref{eq:powerAlloc} based on the MP framework.

\vspace{-1mm}
\section{Proposed Solution for Power Control Problem}
\vspace{-1mm}
\subsection{Equivalent Convex-Concave Reformulation of \eqref{eq:powerAlloc}}

The proposed method is developed based on an equivalent convex-concave reformulation of \eqref{eq:powerAlloc}. To this end, it is easy to see that we can equivalently rewrite \eqref{eq:powerAlloc} as 
\begin{equation}
\underset{\mathbf{p}}{\minimize}\ \underset{1\leq l\leq L}{\max}\gamma_{l}^{-1}(\mathbf{p}),\label{eq:maxmin2minmax}
\end{equation}
where 
\begin{align}
\gamma_{l}^{-1}(\mathbf{p}) & =p_{l}^{-1}\Bigl(\sum\nolimits _{i\neq l}^{L}a_{li}p_{i}+\sum\nolimits _{i=1}^{L}b_{li}p_{i}+c_{l}\Bigr),\label{eq:posynomialObj}
\end{align}
where $a_{li}=\frac{\mathbf{q}_{l}\herm\mathbf{g}_{li}\mathbf{g}_{li}\herm\mathbf{q}_{l}}{\mathbf{q}_{l}\herm\mathbf{g}_{ll}\mathbf{g}_{ll}\herm\mathbf{q}_{l}}$,
$b_{ki}=\frac{(1/K)\mathbf{q}_{l}\herm\bar{\mathbf{G}}_{li}\mathbf{q}_{l}}{\mathbf{q}_{l}\herm\mathbf{g}_{ll}\mathbf{g}_{ll}\herm\mathbf{q}_{l}}$,
and $c_{k}=\frac{(1/K)\mathbf{q}_{l}\herm\tilde{\mathbf{G}}_{l}\mathbf{q}_{l}}{\mathbf{q}_{l}\herm\mathbf{g}_{ll}\mathbf{g}_{ll}\herm\mathbf{q}_{l}}$.
We remark that $\gamma_{l}^{-1}(\mathbf{p})$ is non-convex but can be converted to a single convex form as for geometric programming by making a change of variables $\theta_{i}=\log p_{i}$\cite{Bashar2018mixQoS,Bashar2019maxmin}.
In this letter, \emph{we propose the change of variables $\theta_{i}=\log\omega p_{i}$ or $p_{i}=\frac{1}{\omega}e^{\theta_{i}}$} and reformulate \eqref{eq:posynomialObj}
as 
\begin{align}
f_{l}(\boldsymbol{\theta}) & \triangleq\gamma_{l}^{-1}=\omega e^{-\theta_{l}}\Bigl(\sum\nolimits _{i\neq l}^{L}\frac{a_{li}}{\omega}e^{\theta_{i}}+\sum\nolimits _{i=1}^{L}\frac{b_{li}}{\omega}e^{\theta_{i}}+c_{l}\Bigr)\nonumber \\
 & \negthickspace\hspace{-1cm}=\sum\nolimits _{i\neq l}^{L}a_{li}e^{(\mathbf{e}_{i}-\mathbf{e}_{l})\trans\boldsymbol{\theta}}+\sum\nolimits _{i=1}^{L}b_{li}e^{(\mathbf{e}_{i}-\mathbf{e}_{l})\trans\boldsymbol{\theta}}+\bar{c}_{l}e^{-\mathbf{e}_{l}\trans\boldsymbol{\theta}},\label{eq:convexapprox}
\end{align}
where $\bar{c}_{l}=c_{l}\omega$ and $\boldsymbol{\theta}=[\theta_{1};\theta_{2};\ldots;\theta_{L}]$. Now \eqref{eq:maxmin2minmax} is equivalent to 
\begin{equation}
\underset{\boldsymbol{\theta}\in\Theta}{\minimize}\ \big[f(\boldsymbol{\theta})\triangleq\underset{1\leq l\leq L}{\max}f_{l}(\boldsymbol{\theta})\big]\tag{\ensuremath{\mathcal{P}_{3}}},\label{eq:convexpowercontrol}
\end{equation}
where $\Theta\triangleq\{\boldsymbol{\theta}|\boldsymbol{\theta}\leq\bar{\theta}\}$
and $\bar{\theta}=\log(\omega\bar{p})$. Note that $f_{l}(\boldsymbol{\theta})$
is convex and so is \eqref{eq:convexpowercontrol}. In the context of projected gradient methods, the introduction of $\omega$ effectively scales the gradient of $f_{l}(\boldsymbol{\theta})$. We shall numerically demonstrate that a proper value of $\omega$ can speed up the convergence of the proposed algorithm as shown in the next section.

Although \eqref{eq:convexpowercontrol} is convex, solving it efficiently is still challenging since its objective is nonsmooth due to the max operator. A solution to deal with the nonsmoothness of \eqref{eq:convexpowercontrol} is to adopt a smoothing technique as done in \cite{farooq2020APG} but such a method can only produce an approximate solution. To derive an efficient solution, we recall the following equality
\begin{equation}
\underset{1\leq l\leq L}{\max}f_{l}(\boldsymbol{\theta})=\underset{\boldsymbol{\lambda}\in\Delta}{\max}\ \sum\nolimits _{l=1}^{L}\lambda_{l}f_{l}(\boldsymbol{\theta}),
\end{equation}
where $\boldsymbol{\lambda}=[\lambda_{1};\lambda_{2};\ldots,\lambda_{L}]\in\mathbb{R}^{L}$
and $\Delta\triangleq\left\{ \boldsymbol{\lambda\ }|\ \mathbf{1}\trans\boldsymbol{\lambda}=1;\boldsymbol{\lambda}\geq0\right\} $
is the standard simplex. Thus, \eqref{eq:convexpowercontrol} can be equivalently rewritten as 
\begin{equation}
\boxed{\underset{\boldsymbol{\theta}\in\Theta}{\minimize}\ \underset{\boldsymbol{\lambda}\in\Delta}{\maximize}\ \phi(\boldsymbol{\theta},\boldsymbol{\lambda}),}\tag{\ensuremath{\mathcal{P}_{4}}}\label{eq:minmaxreform}
\end{equation}
where $\phi(\boldsymbol{\theta},\boldsymbol{\lambda})=\sum_{l=1}^{L}\lambda_{l}f_{l}(\boldsymbol{\theta})$.
Note that $\phi(\boldsymbol{\theta},\boldsymbol{\lambda})$ is convex w.r.t $\boldsymbol{\theta}$ for a given $\boldsymbol{\lambda}$ and concave (in fact linear) w.r.t $\boldsymbol{\lambda}$ for a given
$\boldsymbol{\theta}$.

\vspace{-1mm}
\subsection{Proposed Mirror Prox Algorithm for Solving \eqref{eq:minmaxreform}}

It is now clear that we can find a saddle point of \eqref{eq:minmaxreform} to solve the power control problem, which indeed motivates the application of the \ac{MP} algorithm in this letter. To explain the idea of the \ac{MP} algorithm, let $\mathbf{z}=(\boldsymbol{\theta},\boldsymbol{\lambda})$
and $\mathbf{F}(\mathbf{z})=[\nabla_{\boldsymbol{\theta}}\phi(\boldsymbol{\theta},\boldsymbol{\lambda}),-\nabla_{\boldsymbol{\lambda}}\phi(\boldsymbol{\theta},\boldsymbol{\lambda})]$ which is a monotone operator associated with \eqref{eq:minmaxreform}, i.e., $\bigl\langle\mathbf{F}(\mathbf{z})-\mathbf{F}(\mathbf{z}^{\prime}),\mathbf{z}-\mathbf{z}^{\prime}\bigr\rangle\geq0$. The idea of the \ac{MP} algorithm is to find a point $\mathbf{z}^{\ast}$ such that $\bigl\langle\mathbf{F}(\mathbf{z}),\mathbf{z}-\mathbf{z}^{\ast}\bigr\rangle\geq0$, which is indeed a saddle point of \eqref{eq:minmaxreform}. The proposed
method is a special case of the \ac{MP} algorithm with the Euclidean setup. The motivation is that the Euclidean projections onto $\Theta$ and $\Delta$ can be done in closed-form. Under the Euclidean setup, the distance between $\mathbf{z}=(\boldsymbol{\theta},\boldsymbol{\lambda})$ and $\mathbf{z}^{\prime}=(\boldsymbol{\theta}^{\prime},\boldsymbol{\lambda}^{\prime})$, denoted by $D(\mathbf{z},\mathbf{z}^{\prime})$, is defined as 
\begin{equation}
D(\mathbf{z},\mathbf{z}^{\prime})=\frac{1}{2}\bigl\Vert\boldsymbol{\theta}-\boldsymbol{\theta}^{\prime}\bigr\Vert^{2}+\frac{1}{2}\bigl\Vert\boldsymbol{\lambda}-\boldsymbol{\lambda}^{\prime}\bigr\Vert^{2}.
\end{equation}
Note that $D(\mathbf{z},\mathbf{z}^{\prime})$ is the sum of individual Euclidean distances.

The step-by-step description of the proposed method for solving \eqref{eq:minmaxreform} is as follows. Let $\mathbf{z}^{n}=(\boldsymbol{\theta}^{n},\boldsymbol{\lambda}^{n})$ be the $n$-th iterate. To obtain the next iterate, we perform the following two \emph{proximal mappings} 
\begin{subequations}
\begin{align}
\hat{\mathbf{z}}^{n} & =\underset{\mathbf{z}\in\Theta\times\Delta}{\arg\min}\ \bigl\langle\mu^{n}\mathbf{F}\bigl(\mathbf{z}^{n}\bigr),\mathbf{z}\bigr\rangle+D(\mathbf{z}^{n},\mathbf{z}),\label{eq:MP:inter}\\
\mathbf{z}^{n+1} & =\underset{\mathbf{z}\in\Theta\times\Delta}{\arg\min}\ \bigl\langle\mu^{n}\mathbf{F}\bigl(\hat{\mathbf{z}}^{n}\bigr),\mathbf{z}\bigr\rangle+D(\mathbf{z}^{n},\mathbf{z}),\label{eq:MP:nextiter}
\end{align}
\end{subequations}
where $\mu^{n}>0$ is a step size which needs to be chosen properly to guarantee the convergence. We remark that the two steps above are similar to a classical gradient-type method since the vector field $\mathbf{F}(\mathbf{z})$ acts as a descent direction. From the current iterate $\mathbf{z}^{n}$, we first move along the gradient of individual variables to obtain an intermediate point $\hat{\mathbf{z}}^{n}=\bigl(\hat{\boldsymbol{\theta}}^{n},\hat{\boldsymbol{\lambda}}^{n}\bigr)$ as shown in \eqref{eq:MP:inter}. The main difference and in fact
the novel idea of the \ac{MP} algorithm are as follows. To obtain the next iterate from $\mathbf{z}^{n}$, we \emph{do not} use the gradients at $\mathbf{z}^{n}$. Instead, $\mathbf{z}^{n+1}$ is obtained using the gradient at the intermediate point $\hat{\mathbf{z}}^{n}$ as given in \eqref{eq:MP:nextiter}.

\noindent \textbf{Intuition.} To gain further insights into the proposed \ac{MP} algorithm, we note that \eqref{eq:MP:inter} implies

\begin{align}
\hat{\boldsymbol{\theta}} & =\underset{\boldsymbol{\theta}\in\Theta}{\arg\min}\ \bigl\langle\mu\nabla_{\boldsymbol{\theta}}\phi(\boldsymbol{\theta}^{n},\boldsymbol{\lambda}^{n}),\boldsymbol{\theta}\bigr\rangle+\frac{1}{2}\left\Vert \boldsymbol{\theta}-\boldsymbol{\theta}^{n}\right\Vert ^{2}\nonumber \\
 & =\mathcal{P}_{\Theta}\bigl(\boldsymbol{\theta}^{n}-\mu^{n}\nabla_{\boldsymbol{\theta}}\phi(\boldsymbol{\theta}^{n},\boldsymbol{\lambda}^{n})\bigr)
\end{align}
and, similarly, 
\begin{equation}
\hat{\boldsymbol{\lambda}}^{n}=\mathcal{P}_{\Delta}\bigl(\boldsymbol{\lambda}^{n}+\mu^{n}\nabla_{\boldsymbol{\lambda}}\phi(\boldsymbol{\theta}^{n},\boldsymbol{\lambda}^{n})\bigr).
\end{equation}
An important remark is in order. Since $\phi(\boldsymbol{\theta},\boldsymbol{\lambda})$ is convex in $\boldsymbol{\theta}$ and concave in $\boldsymbol{\lambda}$,
$-\nabla_{\boldsymbol{\theta}}\phi(\boldsymbol{\theta},\boldsymbol{\lambda})$
and $\nabla_{\boldsymbol{\lambda}}\phi(\boldsymbol{\theta},\boldsymbol{\lambda})$
are the descent and ascent directions, respectively. In this regard, the \ac{MP} algorithm is similar to the gradient descent and ascent algorithm. That is, the \ac{MP} algorithm simultaneously minimizes and maximizes $\phi(\boldsymbol{\theta},\boldsymbol{\lambda})$ to reach a saddle point. However, the gradients at the intermediate point $\bigl(\hat{\boldsymbol{\theta}}^{n},\hat{\boldsymbol{\lambda}}^{n}\bigr)$
are used to obtain $\bigl(\boldsymbol{\theta}^{n+1},\boldsymbol{\lambda}^{n+1}\bigr)$ from $\bigl(\boldsymbol{\theta}^{n},\boldsymbol{\lambda}^{n}\bigr)$.

To obtain a convergent algorithm, the step size in each iteration $n$ needs to satisfy the following condition

\begin{align}
\delta^{n} & \triangleq\mu^{n}\nabla_{\boldsymbol{\theta}}\phi(\hat{\boldsymbol{\theta}}^{n},\hat{\boldsymbol{\lambda}}^{n})(\hat{\boldsymbol{\theta}}^{n}-\boldsymbol{\theta}^{n+1})\nonumber \\
 & \quad-\mu^{n}\nabla_{\boldsymbol{\lambda}}\phi(\hat{\boldsymbol{\theta}}^{n},\hat{\boldsymbol{\lambda}}^{n})(\hat{\boldsymbol{\lambda}}^{n}-\boldsymbol{\lambda}^{n+1})\\
 & \quad-\frac{1}{2}\bigl\Vert\boldsymbol{\theta}^{n+1}-\boldsymbol{\theta}^{n}\bigr\Vert^{2}-\frac{1}{2}\bigl\Vert\boldsymbol{\lambda}^{n+1}-\boldsymbol{\lambda}^{n}\bigr\Vert^{2}\leq0.\nonumber 
\end{align}
Note that the above inequality is true if $\mu^{n}\leq\mathcal{L}^{-1}$, where $\mathcal{L}$ is the Lipschitz constant of $\mathbf{F}(\mathbf{z})$. In practice, we find $\mu^{n}$ by a back tracking line search. In summary, the proposed  \ac{MP} algorithm for solving \eqref{eq:minmaxreform} is outlined in Algorithm \ref{alg:MP}. 
\vspace{-2mm}
\begin{algorithm}[th]
\caption{MP Algorithm for Solving \eqref{eq:minmaxreform}}
\label{alg:MP}

\begin{algorithmic}[1]

\STATE Initialization: $\mu^{0}>0$, $\rho\in(0,1)$

\STATE $\boldsymbol{\theta}^{1}\gets\boldsymbol{\theta}^{\mathrm{initial}}$,
$\boldsymbol{\lambda}^{1}\gets\boldsymbol{\lambda}^{\mathrm{initial}}$

\FOR{ $n=1,2,\ldots$}

\STATE $\mu^{n}=\mu^{n-1}/\rho$

\REPEAT

\STATE $\mu^{n}\gets\mu^{n-1}\times\rho$

\STATE$\hat{\boldsymbol{\theta}}^{n}\gets\mathcal{P}_{\Theta}\bigl(\boldsymbol{\theta}^{n}-\mu^{n}\nabla_{\boldsymbol{\theta}}\phi(\boldsymbol{\theta}^{n},\boldsymbol{\lambda}^{n})\bigr)$

\STATE$\hat{\boldsymbol{\lambda}}^{n}\gets\mathcal{P}_{\Delta}\bigl(\boldsymbol{\lambda}^{n}+\mu^{n}\nabla_{\boldsymbol{\lambda}}\phi(\boldsymbol{\theta}^{n},\boldsymbol{\lambda}^{n})\bigr)$

\STATE$\boldsymbol{\theta}^{n+1}\gets\mathcal{P}_{\Theta}\bigl(\boldsymbol{\theta}^{n}-\mu^{n}\nabla_{\boldsymbol{\theta}}\phi(\hat{\boldsymbol{\theta}}^{n},\hat{\boldsymbol{\lambda}}^{n})\bigr)$

\STATE$\boldsymbol{\lambda}^{n+1}\gets\mathcal{P}_{\Delta}\bigl(\boldsymbol{\lambda}^{n}+\mu^{n}\nabla_{\boldsymbol{\lambda}}\phi(\hat{\boldsymbol{\theta}}^{n},\hat{\boldsymbol{\lambda}}^{n})\bigr)$

\UNTIL{ $\delta^{n}\leq0$ }

\ENDFOR

\STATE $\boldsymbol{\theta}^{\ast}\gets\frac{\sum_{i=1}^{n}\mu^{i}\boldsymbol{\theta}^{i}}{\sum_{i=1}^{n}\mu^{i}}$,
$\boldsymbol{\lambda}^{\ast}\gets\frac{\sum_{i=1}^{n}\mu^{i}\boldsymbol{\lambda}^{i}}{\sum_{i=1}^{n}\mu^{i}}$

\end{algorithmic}
\end{algorithm}

\textbf{Computation of gradients. }To implement Algorithm \ref{alg:MP},
we need to calculate $\nabla_{\boldsymbol{\theta}}\phi(\boldsymbol{\theta},\boldsymbol{\lambda})$
and $\nabla_{\boldsymbol{\lambda}}\phi(\boldsymbol{\theta},\boldsymbol{\lambda})$
which are given in closed form as 
\begin{equation}
\begin{aligned}\nabla_{\boldsymbol{\theta}}\phi(\boldsymbol{\theta},\boldsymbol{\lambda}) & =\sum\nolimits _{l=1}^{L}\lambda_{l}\nabla_{\boldsymbol{\theta}}f_{l}(\boldsymbol{\theta})\\
\nabla_{\boldsymbol{\lambda}}\phi(\boldsymbol{\theta},\boldsymbol{\lambda}) & =[f_{1}(\boldsymbol{\theta});f_{2}(\boldsymbol{\theta});\ldots;f_{L}(\boldsymbol{\theta})],
\end{aligned}
\end{equation}
where, from \eqref{eq:convexapprox}, we have 
\begin{align}
\nabla_{\boldsymbol{\theta}}f_{l}(\boldsymbol{\theta}) & =\sum\nolimits _{i\neq l}^{L}a_{li}e^{(\mathbf{e}_{i}-\mathbf{e}_{l})\trans\boldsymbol{\theta}}\bigl(\mathbf{e}_{i}-\mathbf{e}_{l}\bigr)\nonumber \\
 & +\sum\nolimits _{i=1}^{L}b_{li}e^{(\mathbf{e}_{i}-\mathbf{e}_{l})\trans\boldsymbol{\theta}}\bigl(\mathbf{e}_{i}-\mathbf{e}_{l}\bigr)-\bar{c}_{l}e^{-\mathbf{e}_{l}\trans\boldsymbol{\theta}}\mathbf{e}_{l}.\label{eq:gradInvSINR}
\end{align}

\begin{rem}
As mentioned above, the proposed change of variable $\theta_{i}=\log\omega p_{i}$
is equivalent to scaling the gradient of $f_{l}(\boldsymbol{\theta})$, compared to the standard change of variable $\theta_{i}=\log p_{i}$. This scaling effect is reflected by the term $\bar{c}_{l}$ in the above equation. The main motivation for introducing $\omega$ is to balance the gradients $\nabla_{\boldsymbol{\theta}}\phi(\boldsymbol{\theta},\boldsymbol{\lambda})$
and $\nabla_{\boldsymbol{\lambda}}\phi(\boldsymbol{\theta},\boldsymbol{\lambda})$,
which can accelerate the convergence of Algorithm \ref{alg:MP} as numerically illustrated in the next section.
\end{rem}
\textbf{Projections onto $\Theta$ and $\Delta$.} It is easy to see that $P_{\Theta}(\mathbf{x})$ admits a closed form solution as follows
\begin{equation}
P_{\Theta}(\mathbf{x})=\begin{cases}
x_{k}, & x_{k}\leq\bar{\theta}\\
\bar{\theta}, & x_{k}>\bar{\theta}
\end{cases},k=1,\ldots,L.\label{eq:projPolyhedral}
\end{equation}
Also, the projection onto a standard simplex is given by 
\begin{equation}
P_{\Delta}(\boldsymbol{\lambda})=\bigl[\boldsymbol{\lambda}-\beta\bigr]_{+},
\end{equation}
where $\beta$ is the solution to the equation 
\begin{equation}
\sum\nolimits _{k=1}^{L}\bigl[\lambda_{k}-\beta\bigr]_{+}=1,
\end{equation}
which can be found by bisection. In summary, to solve \eqref{eq:equivProb} we keep alternately computing \eqref{eq:optimalU} and running Algorithm \ref{alg:MP} until convergence.

\vspace{-1mm}
\subsection{Convergence Analysis}

We first provide the convergence analysis of Algorithm \ref{alg:MP}. Let $f^{\ast}$ be the optimal objective of \eqref{eq:convexpowercontrol} and $\tilde{\boldsymbol{\theta}}^{n}=\bigl(\sum_{n=1}^{N}\mu^{n}\bigr)^{-1}\sum_{n=1}^{N}\mu^{n}\boldsymbol{\theta}^{n}$ be the obtained solution after $N$ iterations. Note that $\tilde{\boldsymbol{\theta}}^{n}$ is the weighted average of the iterates $\boldsymbol{\theta}^{n}$ up to iteration $N$. Then it is shown that $f(\tilde{\boldsymbol{\theta}}^{n})-f^{\ast}\leq\frac{1}{\sum_{n=1}^{N}\mu^{n}}\bigl(\Omega+\sum_{n=1}^{N}\delta^{n}\bigr)$, where $\Omega$ is a constant that depends on the distance generating function. By the line search procedure in Algorithm \ref{alg:MP}, we have that $\delta^{n}\leq0$ and $\mu^{n}\geq\mathcal{L}^{-1}$, and thus, $f(\tilde{\boldsymbol{\theta}}^{n})-f^{\ast}\leq\Omega L/N$. That is, Algorithm \ref{alg:MP} can achieve a $\mathcal{O}(1/N)$-rate of convergence. The proof follows closely the steps in \cite[Proposition 6.1]{juditsky2011first}, and thus, is omitted here for brevity.

\vspace{-1mm}
\subsection{\label{subsec:Complexity-Analysis}Complexity Analysis and Comparison}

We now present the per-iteration complexity analysis of Algorithm
\ref{alg:MP}. It is easy to see that $L$ multiplications are required to compute $f_{l}(\boldsymbol{\theta}^{n})$. Therefore, the complexity of finding the objective is $\mathcal{O}(L^{2}).$ Similarly, we can find that the complexity of $\nabla f_{l}(\boldsymbol{\theta})$ is also $\mathcal{O}(L^{2}).$ It is clear from \eqref{eq:projPolyhedral} that $P_{\Theta}(\mathbf{x})$ has complexity of $\mathcal{O}(L)$. For $P_{\Delta}(\boldsymbol{\lambda})$, by sorting the elements of $\boldsymbol{\lambda}$ in the ascending order, the complexity of
the projection onto a simplex from \cite[Theorem 2.2]{chen2011simplex}
is $\mathcal{O}(L)$ complexity in the worst case. In summary, the per-iteration complexity of the proposed algorithm for solving \eqref{eq:powerAlloc} is $\mathcal{O}(L^{2})$. If a bisection search is used with \ac{LP} as done in \cite{Ngo2017cfmm}, the \emph{worst-case} per-iteration complexity is $\mathcal{O}(L^{3.5})$ \cite[Eq. (8.1.6)]{Nesterov:1994}. The complexity of the \ac{GP}-based methods in \cite{Bashar2018mixQoS,Bashar2019maxmin} is $\mathcal{O}(L^{4.5})$ \cite[Sec. 6.3.1]{Nesterov:1994}.

\vspace{-1mm}
\section{Numerical Results}

We evaluate the performance of our proposed method in terms of the achievable rate and the run time. We randomly distribute \acp{AP} and users over a $D\times D$ \si{\km\squared}. Channel coefficients are generated using \eqref{eq:channel}, in which the large-scale fading coefficient between the $m$-th AP and the $l$-th user is modeled as $\zeta_{ml}=\mathrm{PL}_{ml}z_{ml},$ where $\mathrm{PL}_{ml}$ is the corresponding path loss, and $z_{ml}$ represents the log-normal shadowing between the $m$-th \ac{AP} and the $l$-th user with
mean zero and standard deviation $\sigma_{\textrm{sh}}$, respectively. In this paper, we adopt the three-slope path loss model and model parameters as in \cite{Farooq2021}. Noise figure is set to $9$ dB. We assume pilot sequences to be pairwisely orthogonal to avoid the effect of pilot contamination. The lengths of the coherence interval and the uplink training phase are set to $\tau_{c}=200,\tau_{p}=20$, respectively. If not otherwise mentioned, we set $\eta_{p}=0.2$ W and $\bar{p}=0.2$ W. The number of antennas at each AP is $K=1$. 

\begin{figure}[tbh]
\centering\includegraphics[bb=62bp 552bp 291bp 738bp,width=0.87\columnwidth]{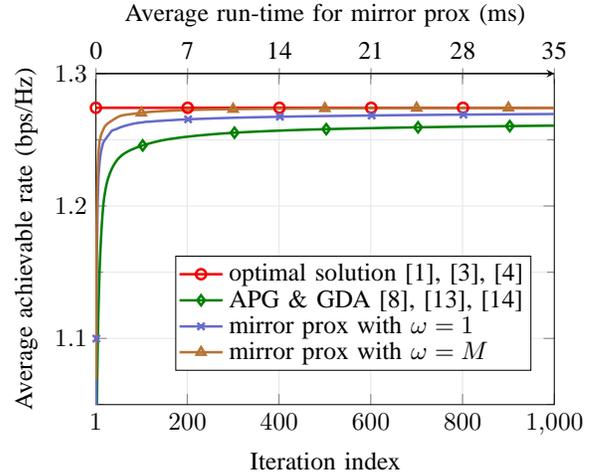}\vspace{-2mm}
\caption{Convergence performance of Algorithm \ref{alg:MP}, \ac{GDA}, \ac{APG}
averaged over $100$ channels. The simulation parameters taken are
$M=150,L=50,$ and $D=1$.}
\label{Fig:convergence}
\end{figure}

In Fig. \ref{Fig:convergence}, we show the convergence of Algorithm \ref{alg:MP}. For each channel realization, we run Algorithm \ref{alg:MP}
for a fixed $\mathbf{q}$ and plot  the average achievable rate over $100$ randomly generated channel realizations. We compare Algorithm \ref{alg:MP} with two other iterative baseline schemes: the \ac{GDA} method \cite{korpelevich1976extragradient,nedic2009subgradient} and the \ac{APG} method combined with the smoothing technique \cite{farooq2020APG}. We also benchmark Algorithm \ref{alg:MP} with the optimal solution obtained by either the bisection method with \ac{LP} \cite{Ngo2017cfmm} or the GP in \cite{Bashar2019maxmin,Bashar2018mixQoS}.

We can see that Algorithm \ref{alg:MP} based on the \ac{MP} method has better objective than \ac{GDA} and \ac{APG}. Also, Algorithm \ref{alg:MP} takes lesser number of iterations. Thus, \ac{GDA} and \ac{APG} \cite{farooq2020APG} are not suitable for the applications where a fast convergence rate is required. Note that \ac{APG} presented in \cite{farooq2020APG} is essentially an approximate solution. On the other hand, \emph{Algorithm \ref{alg:MP} is an exact method and
thus can reach the optimal solution at convergence as clearly seen in Fig. \ref{Fig:convergence}}. Note that we also demonstrate the impact of introducing $\omega$ when reformulating \eqref{eq:powerAlloc} into \eqref{eq:minmaxreform}. As can be observed clearly, a proper value of $\omega$ can indeed speed up the convergence of Algorithm \ref{alg:MP} very significantly. By extensive simulation settings, we find that $\omega=M$ yields a good convergence rate for Algorithm \ref{alg:MP} overall, which is shown in Fig. \ref{Fig:convergence}. Further, we remark that the proposed algorithm converges within $500$ iterations. Also,  the average run-time per iteration is  very small, as shown in Fig. \ref{Fig:convergence}.

\begin{figure}[tbh]
\centering\includegraphics[bb=62bp 552bp 291bp 738bp,width=0.87\columnwidth]{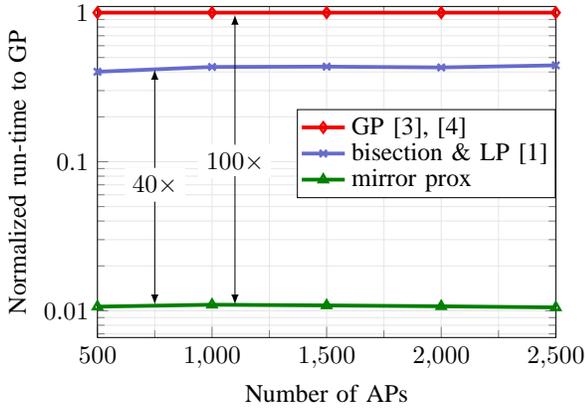}\vspace{-10mm}
\caption{Comparison of run-time (in seconds) between Algorithm \ref{alg:MP}
and the bisection-based method for $L=40$.}
\label{Fig:runtime}
\vspace{-5mm}
\end{figure}

The main advantage of Algorithm \ref{alg:MP} over second-order methods in \cite{Ngo2017cfmm,Bashar2019maxmin,Bashar2018mixQoS} is that each iteration of Algorithm \ref{alg:MP} is very memory efficient and computationally cheap, and hence, can be executed very fast. To demonstrate this, we compare the run-time of Algorithm \ref{alg:MP} with the bisection method with \ac{LP} in \cite{Ngo2017cfmm}, \emph{both normalized to the run-time of the \ac{GP} approach} in \cite{Bashar2018mixQoS,Bashar2019maxmin} for solving
\eqref{eq:powerAlloc} in Fig. \ref{Fig:runtime}. This shows the factor at which a particular method is faster than the GP method with normalized run-time of 1. We run the codes on a $64$-bit Windows operating system with $16$ GB RAM and Intel CORE i$7$, $3.7$ GHz. All iterative methods are terminated when the difference of the objective for the latest two iterations is less than $10^{-4}$. We observe that solving \eqref{eq:powerAlloc} using a bisection search with LP is faster than converting it to a single GP, which is then solved by a dedicated GP solver. Most importantly, Algorithm \ref{alg:MP} is 40 times faster than the LP method and approximately 100 times faster than the GP approach  as shown in Fig. \ref{Fig:runtime}. 

\begin{figure}[tbh]
\centering\includegraphics[bb=62bp 552bp 291bp 738bp,width=0.87\columnwidth]{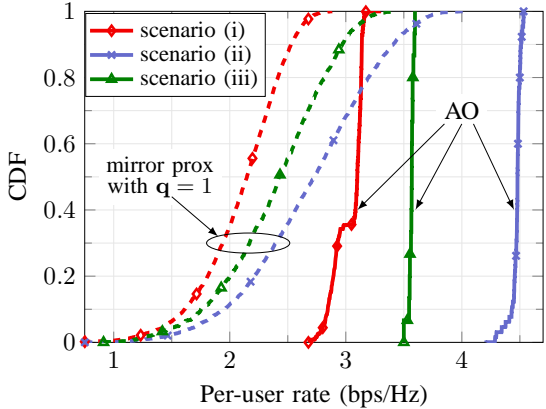}\vspace{-10mm}
\caption{\ac{CDF} of per-user achievable rate for three scenarios: (i)
$M=500,L=50,D=1$, (ii) $M=2000,L=100,D=2$, and (iii) $M=2000,L=50,D=2$. Solid
lines indicate the \ac{CDF} of the per-user rate using both power
 and receiver filtering control, and dashed lines represent the
\ac{CDF} of the per-user rate using power control only.}
\label{Fig:CDFplot}
\vspace{-5mm}
\end{figure}

In the next experiment, we investigate cell-free massive MIMO for large-scale scenarios. Instead of solving a generalized eigenvalue problem to find $\mathbf{q}_{l}^{*}$ as suggested in \cite{Bashar2019maxmin,Bashar2018mixQoS}
which incurs a complexity of $O(M^{3})$, we can reduce this complexity to $O(M^{2})$ using the method in \cite[Appendix]{farooq2020APG}. This complexity reduction and the low complexity of Algorithm \ref{alg:MP} indeed allow us to investigate the performance of uplink large-scale cell-free massive MIMO, which has never been reported previously. In particular, we plot the \ac{CDF} of the per-user achievable rate for $100$ channel realizations in Fig. \ref{Fig:CDFplot}. Three large-scale scenarios detailed in the caption of Fig. \ref{Fig:CDFplot} are considered. Note that the \ac{AP}-density, defined as the number of \acp{AP} per \SI{}{\km\squared} of area, is the same for all three scenarios. In particular, we compare the \ac{CDF} of the per-user rate where both receiver filtering  and power control are considered (i.e. alternately computing \eqref{eq:optimalU} and running Algorithm \ref{alg:MP} until convergence, referred to as AO in the figure) to that where the power control only scheme (i.e. Algorithm \ref{alg:MP}) is adopted in which \ac{AP}-weighting coefficients are all set to one. 

It can be seen from Fig. \ref{Fig:CDFplot} that the system performance improves when the number of APs increases (even the \ac{AP}-density is fixed). Let us discuss scenarios (i) and (iii) first. Although the \ac{AP} density is the same for both scenarios, the user-density (defined as the number of users per \SI{}{\km\squared} of area) for scenario (iii) is much smaller than that for scenario (i), and hence, users in scenario (iii) suffer less interference than users in scenario (i) do. Thus, users in scenario (iii) have higher rates than those in scenario (i). If we increase the number of users in scenario (iii) from $50$ to $100$ as in scenario (ii), then the user-density and thereby inter-user interference increase, which decreases the per-user rate accordingly. Furthermore, it is interesting to note that for scenarios (ii) and (iii), considering both power and  receiver filtering control can deliver more universally good services to the users than the power control only scheme. However, this comes at the cost of extra complexity imposed by the computation of the receiver filter coefficients. Therefore, it is important to consider both the receiver coefficient design and power control problems for large-scale scenarios.

\vspace{-1mm}
\section{Conclusion}

In this paper, we have considered the max-min fairness problem for uplink cell-free massive MIMO. As in previous studies, we have decomposed the problem into the receiver filtering design problem and the power control problem. In particular, we have proposed a novel power control scheme based on the \ac{MP} method. The numerical results have demonstrated that the proposed method is superior to other existing methods such as \ac{LP}-based, \ac{GDA}-based, and \ac{APG}-based methods. More specifically, the proposed method achieves the same objective as these conventional schemes but in much lesser time. The proposed scheme has enabled us to investigate the performance of cell-free massive in  large-scale scenarios. We have also numerically shown that the proposed solution provides better fairness among the users compared to the power control only scheme where \ac{AP}-weighting coefficients are all set to one.

\vspace{-1mm}
\section*{Acknowledgment }
This publication has emanated from research supported by a Grant from Science Foundation Ireland under Grant number 17/CDA/4786.

\vspace{-1mm}
\bibliographystyle{IEEEtran}
\bibliography{IEEEabrv,uplink_ref}

\end{document}